\documentclass[aps,twocolumn,groupaddress,eqsecnum,prb]{revtex4-1}
\usepackage{amsmath,amssymb,amsfonts,graphicx,float,enumerate,bm,hyperref,color}
\def\tr{\textrm{tr}}
\newcommand{\beq}{\begin{eqnarray}}\newcommand{\eeq}{\end{eqnarray}}
\newcommand{\bseq}{\begin{subequations}}\newcommand{\eseq}{\end{subequations}}
\newcommand{\bem}{\begin{pmatrix}}\newcommand{\eem}{\end{pmatrix}}
\newcommand{\ben}{\begin{enumerate}}\newcommand{\een}{\end{enumerate}}
\newcommand{\bea}{\begin{align}}\newcommand{\eea}{\end{align}}
\newcommand{\bef}{\begin{figure}}\newcommand{\eef}{\end{figure}}
\newcommand{\br}{\langle}\newcommand{\ke}{\rangle}
\newcommand{\up}{\uparrow}\newcommand{\down}{\downarrow}\newcommand{\ra}{\rightarrow}
\renewcommand{\vec}[1]{{\mathbf{#1}}}
\newcommand{\mc}{\mathcal}
\newcommand{\mb}{\mathbf}
\newcommand{\mbb}{\mathbb}
\newcommand{\etal}{{\it et al.}}
\newcommand{\nonu}{\nonumber\\}
\newcommand{\pa}{\partial}
\newcommand{\sig}{\sigma}

\newcommand{\eps}{\epsilon}
\newcommand{\sla}{\kern-.5em/}      % Feynman slash

\def\f#1#2{\frac{#1}{#2}}

\begin{document}
\title{Towards the Standard Model for Fermi Arcs from a Wilsonian Reduction of the Hubbard Model}
\author{Seungmin Hong and Philip Phillips}
\affiliation{Department of Physics, University of Illinois at Urbana-Champaign}
\date{\today}

\begin{abstract}
Two remarkable features emerge from the exact Wilsonian procedure for integrating out the high-energy scale in the Hubbard model.  At low energies, the number of excitations that couple minimally to the electromagnetic gauge is less than the conserved charge, thereby implying a breakdown of Fermi liquid theory.  In addition, two charge $e$ excitations emerge in the lower band, the standard projected electron and a composite entity (comprised of a hole and a charge $2e$ bosonic field) which give rise to poles and zeros of the single-particle Green function, respectively.  The poles generate spectral weight along an arc centered at $(\pi/2,\pi/2)$ while the zeros kill the spectral intensity on the back-side of the arc.  The result is the Fermi arc structure intrinsic to cuprate phenomenology. The presence of composite excitations also produces a broad incoherent pseudogap feature at the $(\pi,0)$ region of the Brillouin zone, thereby providing a mechanism for the nodal/anti-nodal dichotomy seen in the cuprates. 
\end{abstract}

\maketitle

\section{Introduction}

As revealed by extensive angle-resolved photoemission (ARPES) studies\cite{shen96, dessau96, norman98, KanigelA06, NormanM07, KingP11, YangH11}, lightly doped copper-oxide superconductors (cuprates) in the pseudogap regime possess a band of excitations that only cross the chemical potential once.  Such a single crossing generates a set of coherent or pole-like excitations that ultimately form a truncated Fermi region, termed a Fermi arc, as opposed to the traditional Fermi surface generated by a double crossing.  The coherent excitations, centered around $(0,0)\rightarrow (\pi,\pi)$, traverse the zone diagonal and terminate in the vicinity of $(\pi,0)$ or $(0,\pi)$, thereby giving rise to a nodal/anti-nodal
dichotomy\cite{Deutscher99, LeTacon06, Tanaka06, Kondo07, Boyer07}, the former being ungapped while the latter is gapped. While some ARPES experiments\cite{MengJ09} performed on Bi$_2$Sr$_{2-x}$La$_x$CuO$_{6+\delta}$ (La-Bi2201) revealed closed hole pockets, and hence consistency with the results from quantum oscillation experiments in high magnetic fields\cite{SebastianS08, JaudetC08, leBoeufD07, husseyQO2008, DN07}, this interpretation has been called into question\cite{KingP11}.  King, \etal\cite{KingP11} observed that the closed pockets seen earlier\cite{MengJ09}  are entirely structural in origin as they originate from overlapping superstructure replicas of the main and shadow bands.  Consequently, the preponderance of evidence from ARPES is that the coherent excitations form a disconnected region in momentum space consistent with a single crossing of the chemical potential.

Theoretically, two questions arise.  1) What suppresses the spectral weight on the back-side of the arc?  2) What is the origin of the incoherent excitations or gap at the zone boundary?  A natural candidate to explain the former is that two kinds of excitations populate a doped Mott insulator, one of which has no overlap with the electron.  Such an excitation will appear in the single-electron Green function as a zero rather than a pole and hence will carry no spectral weight.  In this vein, some have proposed neutral composite excitations\cite{YamajiY11} to explain the origin of Fermi arcs.   Alternatively, Fermi arcs have been seen in numerics\cite{BerthodC06, StanescuT06, imada10} on the 2D Hubbard model and have been modeled phenomenologically\cite{YangK06} (hereafter YRZ).  However, a key assumption of the phenomenological account is that the zero-line is fixed at the diamond-shaped Fermi surface of the non-interacting system.  That the diamond-shaped Fermi surface of the non-interacting system constitutes the zero-line of the single-particle Green function is a rigorous mathematical statement\cite{myzeros} only if the underlying Hamiltonian is particle-hole symmetric.  In fact, from the precise condition\cite{dzy2003, myzeros, rosch2007} for the vanishing of the real part of the Green function, maintaining that the zero-line is doping independent requires unphysical assumptions regarding the spectral function. Certainly such a conservation of the zero-line is not borne out by numerics on the Hubbard model\cite{prelovsek2007, prelovsek2008} nor by analytical arguments\cite{myzeros}. In addition, models involving Cooper pairs, fluctuating or otherwise, have been constructed to either yield arcs\cite{NormanM07} or hole pockets\cite{QiY10,harrisonN07}.  Our discussion here, however, will focus entirely on arcs as this seems to what the ARPES experiments are about.

While the physical origin of arcs might not be clear, the mathematics is. Any Green function of the form
\beq\label{Gfunc}
G_{\rm toy}(\omega, \mb k)=\frac{Z}{\omega-\epsilon_{\mb k}+\frac{|\Delta(\mb k)|^2}{\omega-\varepsilon_{\mb k}}}\label{eq1}
\eeq
will do. The similarity with the BCS Green function is only
perfunctory as there is no anomalous component. Aside from having
poles, Eq.~\eqref{Gfunc} has zeros whenever $\varepsilon_{\mb k}=0$
assuming of course the chemical potential corresponds to
$\omega=0$.   While the two dispersing electronic bands, $\epsilon_{\mb k}$ and $\varepsilon_{\mb k}$ in Eq.~\eqref{Gfunc} are not hard to come by, the parameter $\Delta(\mb k)$ is.  It requires some sort of order, fluctuating or otherwise, or a new bosonic degrees of freedom.  While YRZ proposed Eq.~\eqref{Gfunc} phenomenologically, their intuition was based on weak-coupling RPA diagramatics\cite{konik2006} on 2-leg ladder systems.  In the context of an algebraic charge liquid (ACL)\cite{moon2010}, Qi and Sachdev\cite{QiY10} also obtained Eq.~\eqref{Gfunc}.  Thus far, the only system shown to have the properties of the ACL is one with radically different parameters than the cuprates\cite{khodas2010}. 

Our point here is that $\Delta(\mb k)$ arises fundamentally from a new
degree of freedom associated with dynamical spectral weight transfer.
Ideally, it would be advantageous to derive Eq.~\eqref{Gfunc} directly
from the strong-coupling parameter space of the basic model for a
doped Mott insulator, for example the Hubbard model. Such a derivation
has not presented in the literature. Hence, it is this problem that we
address.  Since zeros\cite{dzy2003,myzeros} arise from a cancellation
of the spectral weight in the upper and lower Hubbard bands (hereafter
UHB and LHB), an accurate description of the upper band is required in
a derivation of Eq.~\eqref{Gfunc}. Hence, an attempt to obtain
Eq.~\eqref{Gfunc} from a model that projects out the UHB, for example
the $t-J$ model, is initially a non-starter as this model does not
have zeros of the type required for Eq.~\eqref{Gfunc}.  Nonetheless,
the information regarding the UHB should be correctly encoded into a
theory of the lower band if the UHB is integrated out (rather than
projected out) exactly {\it{ala}} Wilson\cite{wilson1972}.  In this
paper, we show how the method we have recently
developed\cite{LeighR07, ChoyT08a, ChoyT08b} for carrying out the
Wilsonian program for the Hubbard model can be used to  derive
Eq.~\eqref{Gfunc}.  We show explicitly that two types of excitations
emerge, projected electrons (yielding poles in the propagator) and a new bound state that gives rise to
zeros.  The bound state is not made out of the elemental excitations
and hence is orthogonal to an electron (hence the zero) .  It
represents a charge $e$ excitation that originates from the
non-rigidity of the Hubbard bands, in other words, the well documented
dynamical spectral weight transfer\cite{harris,eskes,sawatzky}, the
key fingerprint of the breakdown of the band concept in Mott systems.
Since the mathematics of Fermi arcs requires two kinds of excitations,
one with poles and the other with zeros, we refer to a physical model
that contains both, such as the one presented here, as the standard
model.  More exotic models relying on some type order would fall
outside this framework.

\section{Charge $2e$ boson theory}

\subsection{Preliminaries}

Underlying our toy Green function is a two-pole structure of the form,
\beq\label{twopoles}
G_{\rm toy}(\omega,\vec k)=Z\left(\frac{\cos^2\theta}{\omega-\omega_+}+\frac{\sin^2\theta}{\omega-\omega_-}\right).
\eeq
 Here $\cos^2\theta=(\omega_+-\varepsilon(\vec
k))/(\omega_--\omega_+)$ and $\omega_\pm=\f12(\epsilon(\vec k)+\varepsilon(\vec
k))\pm\sqrt{(\epsilon(\vec k)-\varepsilon({\vec k}))^2+4|\Delta(\vec
  k)|^2}$.  Zeros arise from the interference between the poles at
$\omega_-$ and $\omega_+$.  Any model that admits zeros must have at least this
two-pole structure.  There are two limits of the Hubbard model in
which the zero surface can be calculated exactly.  In the atomic
limit, the zero surface of the exact single-particle Green function
\beq\label{atomic}
G^R(\omega)=\frac{1+x}{\omega-\mu+\frac{U}{2}}+\frac{1-x}{\omega-\mu-\frac{U}{2}},
\eeq
is independent of momentum given by $\omega=\mu$ and $x=0$.  When the hopping is
non-zero, $t\ne 0$, the only limit in which the zero surface can be
calculated exactly is at half-filling and particle-hole symmetry.  In this limit, the zero surface\cite{myzeros} for a nearest-neighbour band structure
is the magnetic Brillouin zone.  Since Fermi arcs are absent from both the
atomic limit and the half-filled system with hopping, it follows
necessarily that Fermi arcs (if they are present at all in the Hubbard
model) arise entirely from the dynamical part of the
spectral weight. 

Dynamical spectral weight transfer represents a concrete example of
more being totally different.  As is evident from Eq. (\ref{atomic}),
the weight of the lower band in the atomic limit is $1+x$.  This
spectral weight has a natural interpretation in terms of electron
states.  There are $2x$ electron addition and $1-x$ electron
removal states. Hence, in the atomic limit, there is a one-to-one
correspondence with the spectral intensity and the number of electron
states in the lower band.
When the hopping is turned on, the spectral intensity increases in the
lower band to $1+x+\alpha$, where
\beq\label{alpha}
\alpha=\frac{2t}{U}\frac{1}{N}\sum_{ij,\sigma} \langle
f^\dagger_{i{\sigma}}f_{j{\sigma}}\rangle+O((t/U)^2)
\eeq
and the $f_{i\sigma}$'s are a rotation of the original fermionic
operators in the Hubbard model such that it is block
diagonal.  The energy of each block is $nU$, $n$ the number of double
occupancies in each block. $\alpha$ is necessarily positive because
any hopping process that creates doubly occupancy decreases the
available spectral weight in the upper band.  Consequently, counting
electrons, fractionalized or otherwise, cannot exhaust the total
number of degrees of freedom in the lower band.  A new degree of
freedom must be present which is distinct and hence orthogonal to
electron quasiparticles.  This degree of freedom will appear as a zero
in the spectral function in the lower band.  It is precisely the
nature of the states that arise from the mixing with the upper band
that we elucidate here.  

Several approaches suggest some kind of
composite excitation mediates Fermi arcs. 
Consider the $SU(2)$ gauge theory of the
t-J model proposed by Wen and Lee\citep{wenlee96} in which the elemental
fields are the appropriate linear combinations of two charge $e$ bosons and two spinons.  A mean-field calculation\citep{wenlee96} of the
coherent spinon-boson Green function reveals that the spectral weight
of the occupied part of the spectrum exceeds $1-x$, acquiring a value
of $1+x/2$ instead.   Wen and Lee\citep{wenlee96} alleviated this problem by
introducing an interaction which recombined the bosons and fermions
back into the elemental fields.  The effect of this interaction with
strength $U$ was to enhance the spectral
weight in the unoccupied part of the
spectrum.  The correct spectral weights were obtained simply by
adjusting the magnitude of $U$. The unnoccupied
part of the spectrum\citep{wenlee96} shows up as a small hole pocket centered roughly
at $(\pi/2,\pi/2)$. The intensity on the back-side of the pocket is
greatly suppressed, thereby leading to a structure not too distinct
from a Fermi arc. Subsequent work on the $U(1)$ formulation of the t-J
model with a phenomenological spinon-holon binding term reached a
similar conclusion\cite{ng2005}.  However, the most extensive calculation in the
gauge theoretic formulations of the Hubbard model reached a rather different
conclusion. Working directly with the Hubbard model,
Imada and colleagues\cite{imadacomp} used a slave-particle
construction with the gauge fluctuations treated at the RPA level and
concluded it is actually  
dynamical spectral weight transfer that leads to a fermi arc structure
and a Green function of the YRZ form.  
Aside from suffering from the lack of a systematic way of treating the
gauge fluctuations, this formulation generates Fermi arcs
from a neutral composite excitation\citep{imadacomp}.  Since neutral
entities cannot couple to the current, it is unclear how such entities
can influence the spectral function. What we demonstrate here without
resorting to a gauge theory is that
dynamical spectral weight transfer mediates Fermi arcs.

\subsection{Exact Results: The conserved charge and the low energy mode}

Dynamical spectral weight transfer has two profound consequences.
First, we show exactly that the conserved charge ($1-x$) is not
exhausted by counting the degrees of freedom minimally coupled to the
electromagnetic gauge.  The remainder are carried by an incoherent
background.  Second, the dynamically transferred degrees of freedom
give rise to zeros in the lower band.  In fact, the physics we find here is analogous  to the finite $U$ charge processes that contribute to the spectral function measured by ARPES in heavy-fermions\cite{Ghaemi2008}.

We first start with the procedure to integrate out the UHB.  Previously, we demonstrated\cite{LeighR07,ChoyT08a,ChoyT08b} that a theory of the lower Hubbard band (hole doping) can be obtained by introducing a new fermionic field $D_i$\cite{LeighR07, ChoyT08a} which represents the excitations in the upper band.  This field has mass $U$ and hence should be integrated out to construct the exact low-energy theory.  As in the derivation of all collective phenomena which dates back to the classic paper of Bohm and Pines\cite{bp1959}, a constraint must be introduced so that when solved, the model in the extended space is equivalent to the starting UV-complete theory.  The new fermionic degree of freedom enters the action in a quadratic fashion and hence the standard fermionic path integral techniques can be applied to integrate the high-energy scale of the UHB.  Since the lower and upper bands are not rigid in the sense that the spectral weights of the two bands are coupled, integrating out the upper band will lead to new degrees of freedom in the lower band.  In Euclidean signature, the Hubbard action in the extended space is
\begin{widetext}
\begin{align}
\mc S_h^{\text{UV}} 
	&= \int_0^\beta d\tau\int d^{2}\theta
		\biggr\{\bar\theta\theta\sum_{i,\sig}(1-n_{i\bar\sig})c_{i\sig}^*\pa_\tau c_{i\sig}
		+\sum_{i}D_{i}^*\pa_\tau D_{i}+ U\sum_{j}D_{j}^* D_{j}\nonu
	&	- t\sum_{i,j,\sig}g_{ij}
		\Big[\bar\theta\theta(1-n_{i\bar\sig})(1-n_{j\bar\sig})c_{i\sig}^* c_{j\sig}
		+ D_{i}^* c_{j\sig}^* c_{i\sig}D_{j}
		+(D_{j}^*\theta c_{i\sig}V_{\sig}c_{j\bar\sig}+{\rm{c.c.}})\Big]
		+s\bar\theta\sum_j\varphi_j^*(D_{j}
		-\theta c_{j\up}c_{j\down})+{\rm{c.c.}}\biggr\},
\end{align}
\end{widetext}
where the matrix $g_{ij}$ selects the relevant neighbors,
$V_\sig=\pm1(\sig=\up,\down)$, the constraint is given by
$\delta(D_i-\theta c_{i\uparrow}c_{i\downarrow})$, $\theta$ is a
Grassmann, $s$ is a constant with units of energy so that $\varphi_i$
is dimensionless and $c_{i\sigma}$ is an electron annihilation operator for
site $i$ with spin $\sigma$.  Because the $\delta$-function constraint
appears exponentiated in the action, an auxiliary field with charge
$2e$, $\varphi_i$ enters the action as a Lagrange multiplier. As a consequence, the field $\varphi$ is not made out of the elemental excitations (thereby distinguishing it from other charge $2e$ scenarios involving pairs of electrons) but rather arises because the UHB and LHB are not rigid bands. In the
action, the first two terms represent the dynamics in the lower and
upper Hubbard bands, respectively, the third term the mass of the $D$
field, the fourth term the hopping in the lower band with matrix
element $t$, the next two the dynamical mixing between the upper and
lower bands and the last term the constraint.  The constant $s$ has
units of energy and is $O(t)$\cite{charge2e3}. It is
straightforward to check that solving the constraint by integrating
out the auxiliary field, $\varphi_i$, followed by an integration over
$D_i$ exactly reduces $S_h^{\text{UV}}$ to the action for the standard Hubbard model.  This is the UV limit of our theory. The advantage of the reformulation above is that it
cleanly associates the physics of the upper band with a fermionic
field $D_i$ which enters the action in a quadratic fashion.  To obtain the IR limit, one simply has to perform the Gaussian integration over the massive field, $D_i$. The result is the low-energy or IR action, $S_h^{\text{IR}}=\int d\tau \mc L_h^{\text{IR}}$, with the associated Lagrangian, 
\beq\label{eq:IR_lg}
\mc L_h^{\text{IR}}
	&=&(1-n_{i\bar\sig})c_{i\sig}^*\pa_\tau c_{i\sig}
		-tg_{ij} (1-n_{i\bar\sig})c_{i\sig}^* c_{j\sig}(1-n_{j\bar\sig})\nonu
	&&	-\left(s\varphi_i-tb_i\right)^*\big(\mc M^{-1}\big)_{ij}\left(s\varphi_j-tb_j\right)\nonu
	&&	-\left(s\varphi_{i}^* c_{i\up}c_{i\down}+\text{c.c.}\right)-\f 1\beta \tr\ln\mc M,
\eeq
where a matrix element of $\mc M$ is given by $\mc M_{ij}=(\pa_\tau+U)\delta_{ij}-tg_{ij}c_{j\sig}^\dag c_{i\sig}$ and $b_i=\sum_jg_{ij}c_{j\sig}V_\sig c_{i\bar\sig}$. Hereafter, repeated indices are implicitly summed unless otherwise stated.  It is important to note that no approximations have been made as of yet. 

In both actions, $S_h^{\text{UV}}$ and $S_h^{\text{IR}}$, global $U(1)$ symmetry guarantees the existence of a conserved charge, which turns out to be
\begin{align}
Q_h^{\text{UV}} &= (1-n_{i\bar\sig})c^*_{i\sig}c_{i\sig}+2D_i^*D_i\label{eq:Q_UV},\\
Q_h^{\text{IR}} &=(1-n_{i\bar\sig})c^*_{i\sig}c_{i\sig}\nonu
	&\quad +2\left(s\varphi_i-t b_i\right)^*\left(\mc M^{-1}\right)_{ik}\left(\mc M^{-1}\right)_{kj}\left(s\varphi_j-t b_j\right)\label{eq:Q_IR}.
\end{align}
It is a natural consequence that the conserved charge $\br Q_h^{\text{UV}}\ke$ is consistent with the number of electrons in the original Hubbard model since the operator $D_i^\dagger D_i$ counts the number of doubly occupied sites. From the Hellman-Feynmann theorem, it is straightforward to check how the number of double occupancies, $n_{\text{docc}}^{\text{UV/IR}}$, is related to $D_i$. Since $n_{\text{docc}}^{\text{UV/IR}} = \beta^{-1}{\pa\ln\mc Z_h^{\text{UV/IR}}}/{\pa U}$ with $\mc Z_h = \int D[\cdots] e^{-\mc S_h}$, one can easily observe the second terms in Eq.~\eqref{eq:Q_UV} and \eqref{eq:Q_IR} are identical to $n_{\text{docc}}^{\text{UV}}$ and $n_{\text{docc}}^{\text{IR}}$, respectively. As a result, the conserved charge $\langle Q_h^{\text{IR}}\rangle$ is identified with the number of electrons, $1-x$, with $x$ the number of holes. This is one of the indications that the low-energy action, $S_h^{\text{IR}}$, retains the structure of the Hubbard model even after the integration of the massive modes.

Another advantage of the low energy theory is that the non-Fermi-liquid nature of the low-energy excitations is immediately  manifest. To illustrate this, one can add a minimally coupled source term\cite{ChoyT08a}
\begin{align}
\mc {L'}_h^{\text{UV}}
	&= J_{i\sig}^*\left[\bar\theta\theta(1-n_{i\bar\sig})c_{i\sig}
	+ \bar\theta c_{i\bar\sig}^*V_\sig D_i\right]+\text{c.c.}
\end{align}
so that when the constraint is solved, the bare electron operator is generated\cite{ChoyT08a}. What we would like to know is what is the transformed fermion at low energies.  Integrating out the $D_i$ fields, results in a source contribution to the IR Lagrangian,
\begin{align}\label{LIRJ}
\mc {L'}_h^{\text{IR}}
	&= J_{i\sig}^* \psi_{i\sig}+\text{c.c.} - J_{i\sig}^* c_{i\bar\sig}^*\left(\mc M^{-1}\right)_{ij}c_{j\bar\sig}J_{j\sig}
\end{align}
with a new collective field, $\psi_{i\sig}$ given by
\bea\label{eq:low_e_mode}
\psi_{i\sig}^* &= (1-n_{i\bar\sig})c_{i\sig}^* +tb_j^*\left(\mc M^{-1}\right)_{ji}V_{\sig}c_{i\bar\sig}\nonu
	&\quad-s\varphi_j^*\left(\mc M^{-1}\right)_{ji}V_{\sig}c_{i\bar\sig}.
\end{align}
Note $\psi_{i\sigma}$ is derived not contrived.  It is Eq.~(9) in the Ref.~35. We obtained it by integrating the $UV-$ complete Lagrangian in the
presence of the source term that generates the correct UV current with
respect to the massive field $D_i$.  $\psi_{i\sigma}$ is the
propagating degree of freedom in the IR. It contains not only an
electron-like quasiparticle affected by nearby spin fluctuations, but
also a hole (with the opposite spin) that is dressed with a
doubly-charged bosonic mode.  Note, we cannot give $\psi_{i\sig}$ a
simple interpretation in terms of bosons or fermions.  At best,
$\psi_{i\sig}$ corresponds to the physical field that is minimally
coupled to an external gauge field. That is, these are the excitations
that couple to light.  Hence, it is the field that is probed by an
ARPES experiment, for example.  While $\psi_{i\sigma}$ was derived
earlier, what we did not show explicitly is that it does not  stand in
a one-to-one correspondence with the bare electrons.  This can proven
exactly by focusing on the positive-definite quantity,
\begin{align}\label{eq:n_psi}
\psi_{i\sig}^*\psi_{i\sig}
	&= (1-n_{i\bar\sig})c_{i\sig}^*c_{i\sig}\nonu
	&\quad +\left(tb-s\varphi\right)_j^*\left(\mc M^{-1}\right)_{ji}c_{i\bar\sig}
				c_{i\bar\sig}^*\left(\mc M^{-1}\right)_{ij}\left(t b-s\varphi\right)_j\nonu
	&= Q_h^{\text{IR}}\nonu
	&\quad -\left(tb-s\varphi\right)_j^*\left(\mc M^{-1}\right)_{ji}(2+c_{i\bar\sig}^*
				c_{i\bar\sig})\left(\mc M^{-1}\right)_{ij}\left(t b-s\varphi\right)_j,
\end{align}
which is essentially the conserved charge less the number of doubly
occupied sites.  Since the second term in the last line is positive definite, the
number of low-energy collective modes which are minimally coupled to
the electromagnetic gauge field is less than $\langle Q_h^{\rm IR}\rangle=
1-x$.  The natural resolution of this conumdrum is that the
number operator only counts those excitations that have a
particle-like interpretation. That is, the number
operator only counts the coherent part of the spectrum.  All of the
stuff mediated by mixing with the upper band is entirely incoherent
and hence while it can contribute to the current, it is not enumerated by
counting the number of particles.  The remainder of the charge count is carried
by the last term in Eq.~\eqref{LIRJ}.  

This discrepancy is not a
surprise when one considers that the total spectral weight of the
lower band exceeds $1+x$\cite{harris,sawatzky} by a dynamical
correction, $\alpha>0$, that depends on the hopping integral, $t$.
Since there are only $1+x$ electron states in the lower band, and only
charge $e$ excitations contribute to the spectral function, there has
to be some new charge mode to make up the difference.  What
$\psi_{i\sig}$ lays plain is that there are charge $e$ states that
contribute to the current that are completely incoherent.  It is a
composite excitation of $\varphi^\dagger$ and a hole $c_{i\bar\sig}$.
In terms of the UV variables, this degree of freedom represents the
binding of a doublon and a holon.  The new composite excitation,
$\varphi^*\mc M^{-1}V_\sig c_{\bar\sig}$ has internal structure and
hence is orthogonal to the projected electron. Since there is no
Hilbert space for $\varphi$, interpreting $\varphi^\dagger
c_{i\bar\sigma}$ in terms of a particle is not possible.  It is this additional degree of freedom that creates the Fermi arc structure--that is, the zeros of of the Green function.  Hence, hidden in $\psi_{i\sig}$ is an incoherent contribution to the single-particle Green function.  What this discussion makes clear is that $\varphi$ should not be considered to be an independent degree of freedom but rather one that is strongly coupled to the fermions. 

What we have shown thus far is that there is a dynamical contribution
to the charge degrees of freedom that are coupled to the source term
that generates the current.   Such entities are the
physical degrees of freedom that create holes in the lower band.
Consequently, when one such excitation is removed from the lower
band, the change in the spectral weight should also depend on $t$.
Hence, the doping level should receive a dynamical contribution.  To
this end, we defined\cite{shila2010} $x'=x+\alpha$ and hence the weight in the UHB is
$1-x'$ and the occupied and empty parts of the lower band are $1-x'$
and $2x'$, respectively.

\subsection{Green function and the approximations}

Thus far, all of our statements are exact.  Our calculation of the
Green function is not, however.  To lend credence to our treatment, we
state our assumptions clearly and up front. The complexity arises in
treating the $\varphi$ degree of freedom.  Our treatment is in the
spirit of the results obtained in the previous section, namely that $\varphi$
leads to the creation of a new charge $e$ excitation that is
orthogonal to a projected electron on account of its internal
structure. 

Having determined the generating functional, ${\mc L'}_h[\{J_{i\sig}^*,J_{i\sig}\}]$, we proceed to calculate the Green function. In the functional formalism, it is given by
\begin{align}
&G_\psi(\mb r_i-\mb r_j,\tau) \nonu
&= -\left.\f{\delta^2}{\delta J_{i\sig}\delta J_{j\sig}^*}\ln \mc Z_h^{\text{IR}}[\{J_{i\sig}^*,J_{i\sig}\}]\right|_{J^*=J=0}\nonu
&= -\left\br\mc T_\tau \psi_i(\tau)\psi_j^\dag(0)\right\ke+\left\br\delta(\tau)c_{i\bar\sig}^\dag\left(\mc M^{-1}\right)_{ij}c_{j\bar\sig}\right\ke,
\end{align}
where $\mc T_\tau$ represents time ordering and $\br\cdots\ke$ stands
for the average over all possible paths. Since the second term is
independent of time, this term contributes to the incoherent part of
the Green function. To understand the first term which contains both
coherent as well as incoherent responses, it is sufficient to focus on the correlator
between the $\psi_{i\sig}$'s. Since $\psi_{i\sig}$ contains a
composite excitation which has a prefactor of $t/U$, $\br \mc T_\tau
\psi_{i\sig}(\tau)\psi_{j\sig}^\dag(0)\ke$ can, in principle, be
expanded in power of $t/U$. The presence of the projection operator,
$(1-n_{i\bar\sig})$, however, does not
 necessarily guarantee that it gives a dominant contribution compared to the composite entities.

For the purpose of numerical evaluation, we make an approximation to
the projection operators, following the idea developed by Zhang, \etal
~\cite{ZhangF88} However, the crucial difference is that we make
the substitution the bare hole concentration by the effective hole doping 
level ($x\rightarrow x'$), since the physical entities
coupled to the external gauge field are not the bare electrons but are
rather dynamically generated. Hence our first approximation is
\ben[]
\item ({\bf A-1})\quad
	$(1-n_{i\bar\sig})c^*_{i\sig}c_{j\sig}(1-n_{j\bar\sig})\rightarrow g_tc^*_{i\sig}c_{j\sig}$,
\een
where $g_t=2x'/(1+x')$.   Interestingly, in
the strong coupling limit, ($U/t\gg1$), a mean-field approach to
Kotliar-Ruckenstein's slave boson construction\cite{KotliarG86} led to
the same renormalization factor for the charged fermion but with $x'$
replaced with $x$. Likewise, 
\ben[]
\item ({\bf A-2})\quad
	$(1-n_{i\bar\sig})c_{i\sig}^\dag (\pa_\tau+\cdots) c_{i\sig}\rightarrow g_pc_{i\sig}^\dag (\pa_\tau+\cdots) c_{i\sig}$.
\een
where $g_p=(1-x')/(1-x)$. The multiplicative factors are chosen here for
internal consistency with the two assumptions.

Since the action $\mc S_h^{\text{IR}}$ has all relevant degrees of
freedom for the low-energy sector, including the spin singlet
fluctuations ($b_i$) and mixing between the separate Hubbard bands
($\varphi$), it is reasonable to expand the action in powers of
$t/U$. To leading order, the matrix elements $(\mc M^{-1})_{ij}$ is
$U^{-1}\delta_{ij}$. From the fact that the collective boson
$\varphi_i$ only has dynamics through its coupling to the fermions, we
assume the dynamics of the boson to be frozen.  Operationally this
assumption breaks down at $O(t/U)^2$ where the explicit dynamics of
$\varphi$ appears as can be seen from an expansion of the ${\cal M}$
matrix,
$\f{s^2}{U^2}\varphi^*(\pa_\tau-U+\cdots)\varphi$. In fact, this even
at $O(t/U)^2$, the propagator for $\varphi$ lays plain that it has a
pole only in the high-energy sector.  This justifies the
assumption that 
\ben[]
\item ({\bf A-3})\quad Bosonic field, $\varphi$, has no dynamics in the LHB.
\een
In other words, it alone is highly massive and is not likely to
propagate in the low-energy sector. Finally, although local spin ordering
might non-negligible, we will assume it to be at most ancillary to
the strong interaction physics arising from the coupled boson-fermion
terms. This is a key assumption and certainly not traditional as most
treatments of the LHB focus on the spin physics. However, as our
emphasis here is on isolating the source of zeros in the LHB,
demonstrating that the action possesses such modes in the absence of
the spin-spin scattering term would suffice to show that such an interaction is
indeed ancillary to the essential charge physics.   As will become evident, our treatment does in fact show
this to be the case.   
Under these considerations, the effective low energy action turns into
\begin{align}\label{eq:Apprx_S}
\mc S_h^{\text{IR}}
	=& \int_0^\beta d\tau\biggl\{
		c_{\mb k\sig}^*\big[g_p(\pa_\tau-\mu)\delta_{ij}-g_t\eps_{\mb k}\big] c_{\mb k\sig} \nonu
	& -\f1{U^2}\left(s\varphi-tb\right)_{\mb q}^*(U+2\mu)\left(s\varphi-tb\right)_{\mb q}\nonu
	& -(s\varphi_{\mb q}^*c_{\mb q-\mb k\up}c_{\mb k\down}+\text{c.c.})\biggl\},
\end{align}
where $\mu$ denotes the chemical potential, $\mb k$ and $\mb q$ are
the momenta, and $\eps_{\mb k}=t g_{ij} e^{i\mb k\cdot(\mb r_j-\mb
  r_i)}$.  This action has a BCS-like coupling and hence will have a
Green function of the form of Eq.~\eqref{eq1}.  That the Green function must
be of the form of Eq.~\eqref{eq1} is not dependent on the assumptions
delineated earlier.  It relies solely on the fact that the spectral
weight in the lower band exceeds $1+x$ and hence a new charge $e$
state distinct from the projected electrons must be present.  Such an
excitation can only be a composite.

Consequently, for a given amplitude of $\varphi_{\mb q}$, the Fourier transformation of the two point correlator $\mc G=-\br\mc T_\tau \psi_i(\tau)\psi_j^\dag(0)\ke$ becomes

\begin{subequations}
\begin{align}
\mc G(i\omega_n,\mb k) &=\f{\widetilde{g_t}}{i\omega_n-\mu-\widetilde{g_t}\eps_{\mb k}-\Sigma_\pm(i\omega_n,\mb k)}+\f tU(\cdots),\label{GTU}\\
\Sigma_\pm(i\omega_n,\mb k) &= \f{s_{\mb k,\mb q}^2\varphi_{\mb q}\varphi_{\mb q}^*}{i\omega_n-\mu\pm\widetilde{g_t}\eps_{\mb q-\mb k}},
\end{align}
\end{subequations}
where $\omega_n=(2n+1)\pi/\beta$ for $n\in\mbb Z$, $\widetilde{g_t}=g_t/g_p$, and
$s_{\mb k,\mb q}=1-(\eps_{\mb k}+\eps_{\mb q-\mb k})/U$. The $\pm$
subscript on the self energy arises from the two choices which are
possible to treat the dynamics of the charge $2e$ boson.  If
$\varphi_i$ is treated as an independent degree of freedom that can
condense, then it can be absorbed as a redefinition of the interaction
strength, $s\rightarrow s\varphi$.  This will correspond to a simple
condensation of $\varphi$ in a non-zero momentum particle-particle
channel, hence the $+$ sign in front of the
$\tilde g_t\epsilon_{\mb q-\mb k}$ factor in the denominator of the self energy.  As will be
clear, this is not the interpretation of $\varphi$ that is ultimately
consistent with the theory outlined here.  Alternatively,
$\varphi_i^\ast c_{i\bar\sigma}$ could be viewed as a new composite
charge $e$ excitation that results from dynamical spectral weight
transfer.   With such bound modes, the interaction term, $\varphi^* c_{\up} c_{\down}$, can
be interpreted as a particle-hole scattering process.  To implement
this interpretation in the Green function, we note that since
$\varphi^* c_{\up} c_{\down}$ now describes the scattering
of an electron $c_\sig$ off a composite particle $\varphi^* V_\sig c_{\bar\sig}/|\varphi|$, the denominator in the
one-loop self energy will resemble that of a particle-hole scattering
event, thereby leading to a $-$ sign in front of the $\tilde g_t\epsilon_{\mb q-\mb k}$ term
in the denominator of the self energy.  In additon, the ellipse
in Eq.~\eqref{GTU} represents the terms that originate from the
mixing between the composite excitations and the projected electron,
which are at least suppressed by the factor $t/U$. The number
$\widetilde{g_t}$ results from the rescaling $g_p^{1/2}c_{i\sig}\ra
c_{i\sig}$. Since $\widetilde{g_t}=g_t/g_p\simeq 2x'$, the $t/U$
corrections in Eq.~\eqref{GTU}  are comparable to the leading term
only for $x'<t/2U\sim 0.05$.  For example, at half-filling, the Green
function only has the $t/U$ term and the spectral weight is governed
entirely by the mixing between the projected and composite excitations
as shown previously\cite{charge2e3}. In the current treatment, we will
explore entirely the contribution from the leading term which is of
the form of Eq.~\eqref{eq1}.

\subsection{Free-energy Minimum approach to $2e$ boson}

Evaluating the Green function is equivalent to a random-matrix problem.  
In the most general case, the field $\varphi_i$ must be integrated over with a
separate value on each site.  However, such
a multi-variable integration is not tractable in any dimension.   From the observation that the collective
boson is not canonical, that is, it does not have its own kinetics, it
was previously conjectured that the spatially homogeneous
configuration was the most prominent candidate for the ground
state\cite{LeighR07,charge2e3}. Even though such an
approach was successful in capturing some experimental\cite{charge2e3}
findings, it still leaves an open question whether the homogeneous
solution minimizes the free energy.  To this end, we explore some
inhomogeneous solutions for $\varphi_i$ to see where the free-energy
is a minimum.  In particular, we explore a staggered configuration,
$\varphi_i=e^{i\vec q\cdot\vec r_i}|\varphi_0|$.  It should be noted 
that a particular choice of the configuration of $\varphi$ does not correspond to spontaneous 
symmetry breaking, since the bosonic mode is in lack of inherent
dynamics.

\begin{figure}[!t]
\includegraphics[width=0.95\columnwidth]{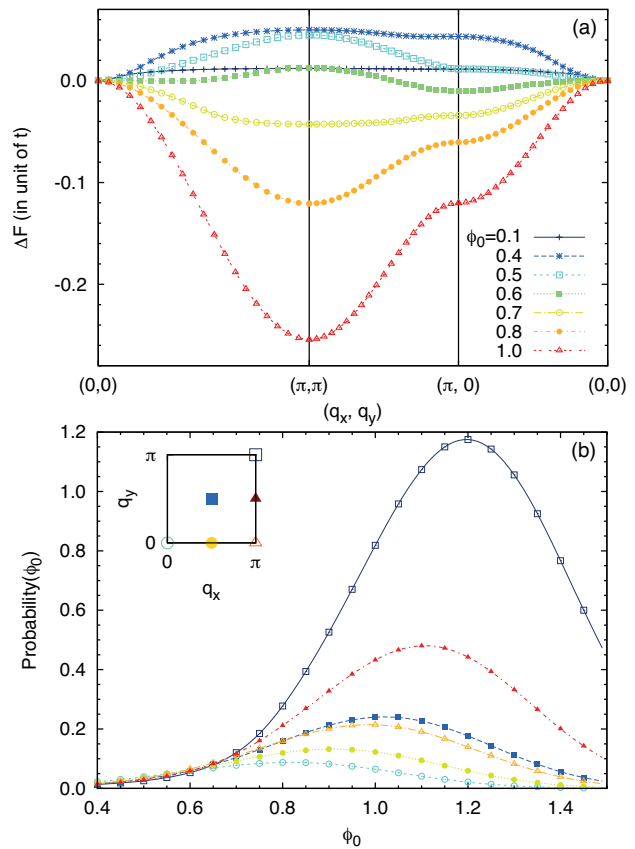}
\caption{(Color online) Free energy minimization and probability
  distribution of charge $2e$ boson.  (a) Plotted here is the $\Delta F=F(\varphi_0
  e^{i\vec q\cdot \vec r})-F(\varphi_0)$ as a function of momentum $\vec
  q$.  $\varphi_0$ is the magnitude of the charge $2e$ boson. As is
  evident, the free energy is strongly dependent on momentum.  As
  $\varphi_0$ increases, the minimum of the free energy shifts to
  $(\pi,\pi)$.  This corresponds to a sign change for the bosonic
  excitation around a plaquette.  (b) Shown here is the probability
  distribution for $\varphi_0$  evaluated from the IR action by the relation, $P(|\varphi|)=1/\mc Z_h^{\text{IR}}\int D[\{c,c^*\}]e^{-\mc S_h^{\text{IR}}}$.  The maximum occurs at
  $\varphi_0\approx 1.2$ where the corresponding momentum that
  minimizes the free energy is $\mb q=(\pi,\pi)$.  }\label{fig:F_energy}
\end{figure}

In Fig.~\ref{fig:F_energy}, we directly compute the free energy difference between a configuration with a spatial texture and the homogeneous state, $\Delta F=F(\varphi_0e^{i\vec q\cdot \vec r})-F(\varphi_0)$, where $\vec q$,
determines the spatial dependence of $\varphi_i$. Except for small
values of $\varphi_0$ in which the homogeneous solution minimizes the
free energy, the distinct minimum occurs at $(\pi,\pi)$ when the
magnitude of the bosonic field,
$\varphi_0$, approaches unity.  This is significant because the
probability distribution of $\varphi_0$ computed from
$P(|\varphi|)=1/\mc Z_h^{\text{IR}}\int D[\{c,c^*\}]e^{-\mc
  S_h^{\text{IR}}}$, has a distinct maximum precisely at the value of
$\varphi$ where the $(\pi,\pi)$ solution minimizes the free energy.
This state of affairs obtains because a quick
inspection of the action reveals that for a staggered configuration of
$\varphi$, the $\varphi^\dagger b$ term actually vanishes.
This results in a lowering of the energy relative to the homogeneous
solution.

\begin{figure}[!t]
\includegraphics[width=3.3in]{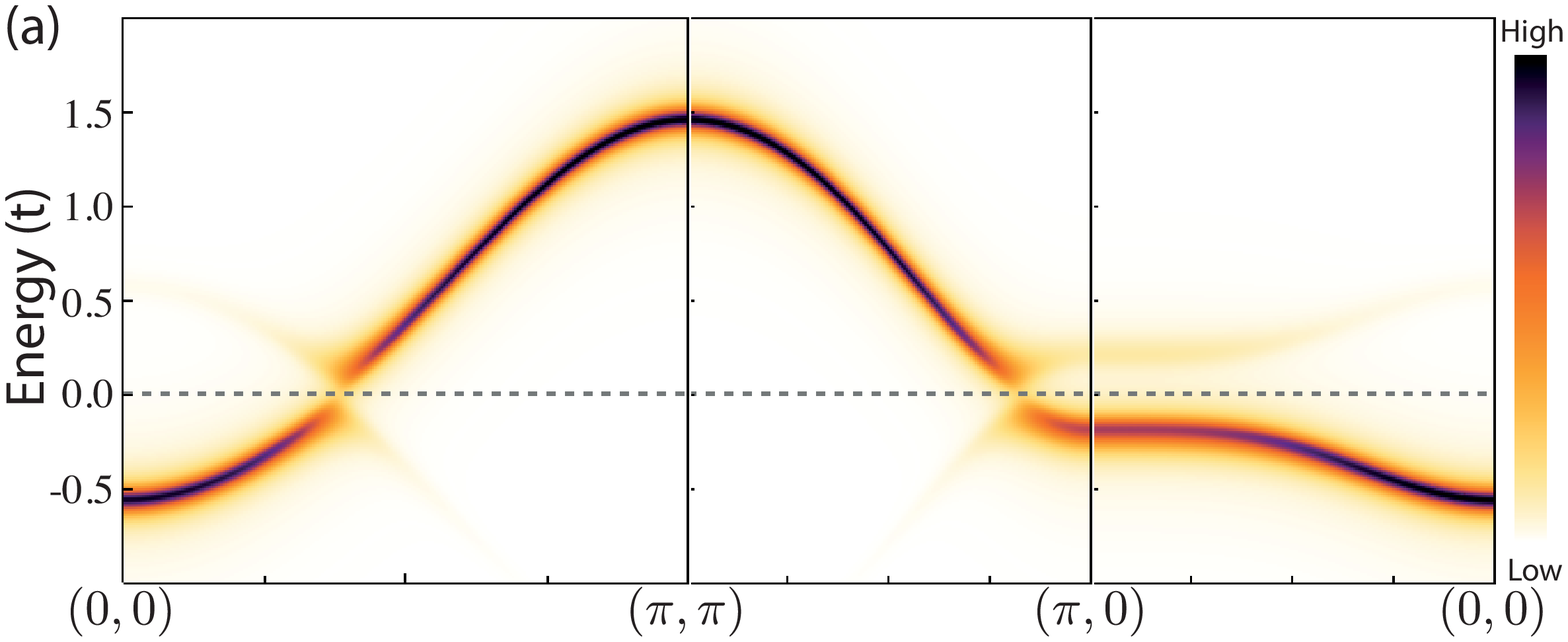}\\
\includegraphics[width=3.3in]{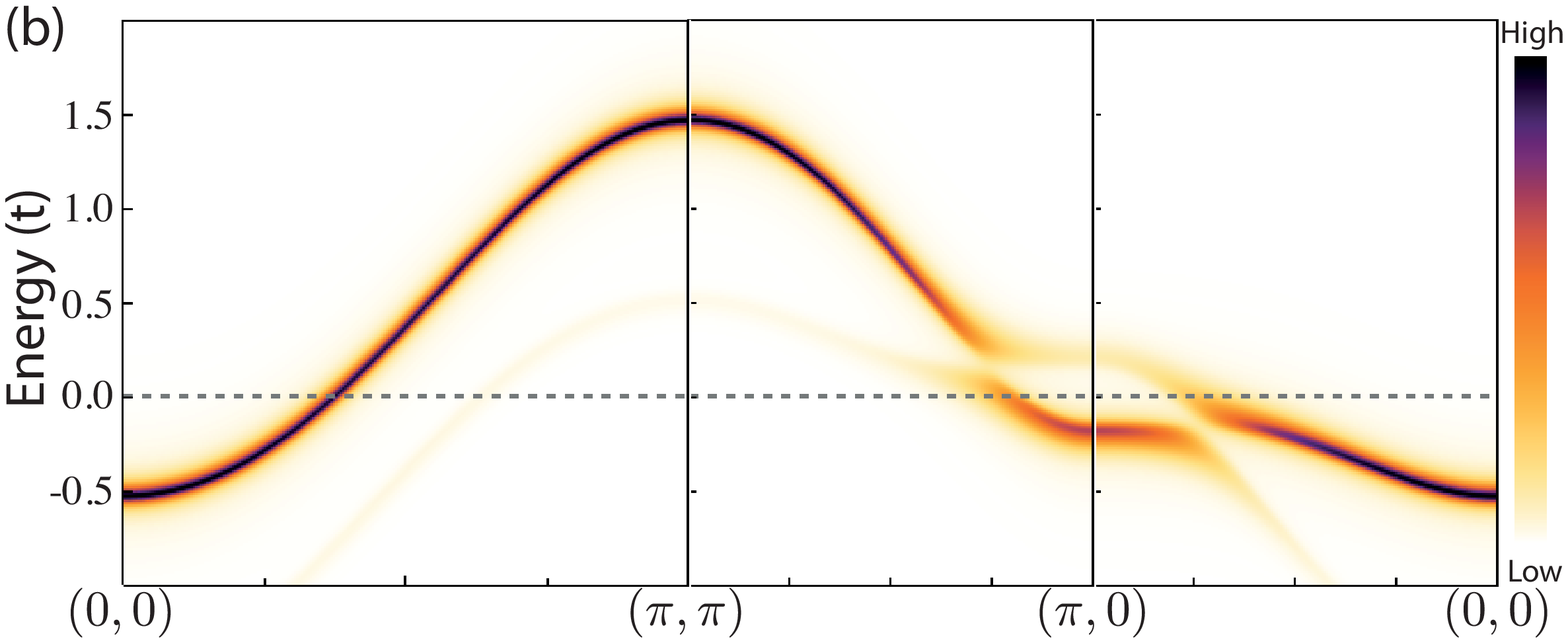}\\
\includegraphics[width=3.3in]{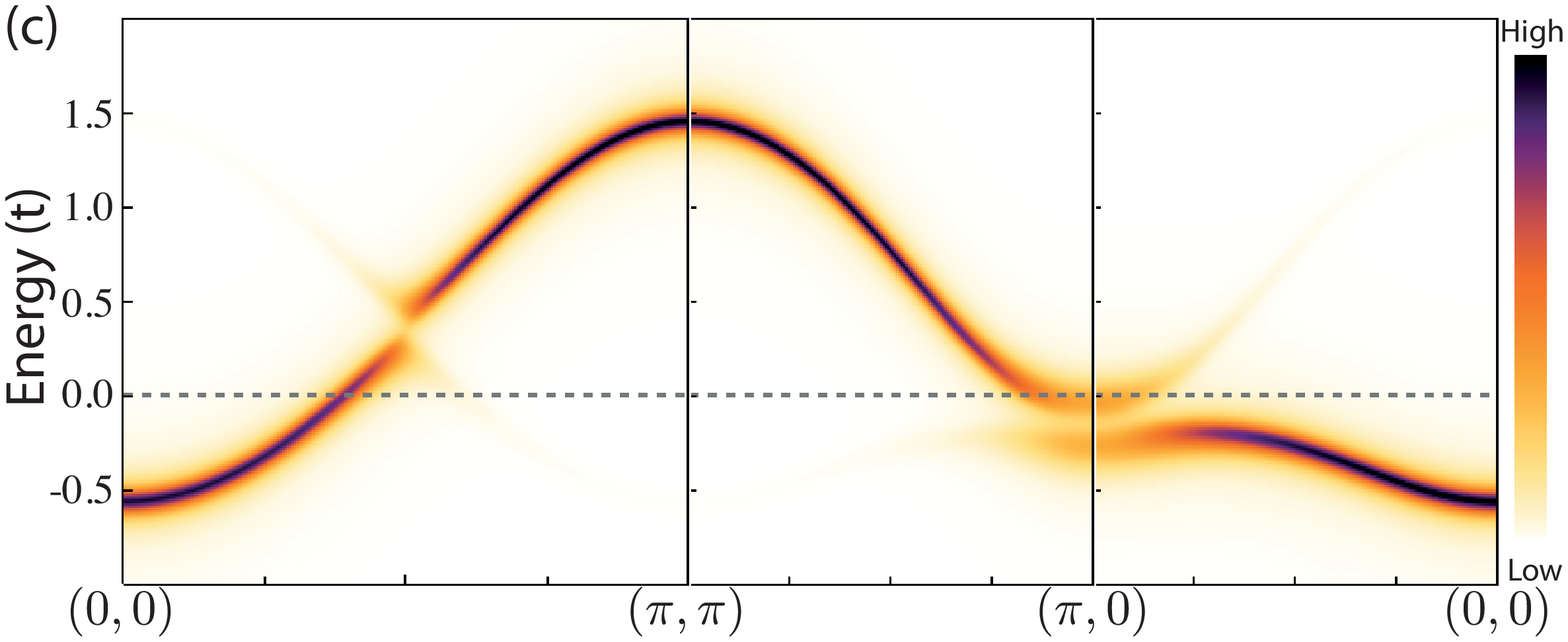}
\caption{(Color online) The low energy band dispersion along high symmetry directions for (a) a homogeneous configuration, $\varphi_i=\varphi_0$, (b) a staggered configuration ($\varphi_i=|\varphi_0|e^{\i\bm\pi\cdot\mb r_i}$) evaluated by Eq.~\eqref{GTU}, and (c) a staggered one evaluated by Eq.~\eqref{GTU}. Here, we take the broadening factor $\eta=0.025t$ for a typical value $U/t=10$, $t'/t=-0.3$, $t''/t=0.1$, and the bare hole doping level as $x=0.12$. In the evaluation, the parameters, $\alpha=x'-x$ are taken from the numerical estimates of the number of double occupancies from Liebsch and Tong\cite{LiebschA09}.}\label{fg:disp}
\end{figure}

That the $(\pi,\pi)$ configuration of $\varphi_i$ minimizes the free
energy is highly significant because the evaluation any integral over
$\varphi_i$ will be dominated by the staggered solution. 
What about the single-particle Green function?  In our previous
treatment of this problem in which we assumed that the mixing with the
UHB was mediated by a homogeneous boson, $\varphi_0$ for all sites, we
obtained a completely gapped structure at the chemical potential for
the spectral function.  Given that $\vec q=(\pi,
\pi)$ is the blobal minimum, our expression for the Green function
simplifies to
\beq
G(i\omega_n,\vec k)=\int d|\varphi||\varphi| P(\varphi){\cal
  G}(i\omega_n,\vec k)|_{\varphi_{\vec q}=\delta_{\vec
    q,\bm\pi}|\varphi|}.\nonumber\\
\eeq
The probability distribution function $P(\varphi)$ is shown in Fig.~1b.
For completeness, we present in Fig.~\ref{fg:disp} the band dispersion
corresponding  to the maximum in the spectral function obtained from
Eq.~\eqref{GTU}  for three different cases: 1) homogeneous solution,
2) staggered $(\pi,\pi)$ phase of $\varphi_i$ in the particle-particle
channel, $\Sigma_+$ and 3) staggered solution in particle-hole
channel, $\Sigma_-$. For the homogeneous phase, we find a hard gap,
Fig.~\ref{fg:disp}(a), because no momentum states cross the chemical
potential. However, as shown earlier, the homogeneous solution does
not correspond to a minimum in the free energy.  Consider the
staggered solutions shown in panels Fig.~\ref{fg:disp}(b) and Fig.~\ref{fg:disp}(c).  
Fig.~\ref{fg:disp}(b) shows that even a staggered solution in the
particle-particle channel, a gap
does not occur at the $(\pi,0)$ region of the Brillouin zone.  There
is also a crossing along the zone diagonal.   This indicates that a simple condensation of
$\varphi$ in a non-zero momentum particle-particle channel cannot give rise to the
nodal/anti-nodal dichotomy.   There is in fact a clear reason why
$\varphi$ cannot be treated as an independent degree of freedom that
can condense.  There is a one-to-one correspondence between 
Eq.~(\ref{eq:low_e_mode}) and its analogue (Eq.~(19) of Ref.\cite{eskes}) in the standard perturbative treatment
of the Hubbard model.  In essence, the charge
$2e$ boson replaces a string of operators that account for the
mixing of double occupancy into the lower band. This is why this
approach is simpler. As it would be
completely incorrect to replace that string of operators with an
average value, it is equally wrong to treat $\varphi$ as a variable
that can condense.  In fact, it is well known\cite{sawatzky} that such mean-field truncations
fail to describe dynamical spectral weight transfer in the Hubbard
model.  

\begin{figure}[!t]
\includegraphics[width=3.3in]{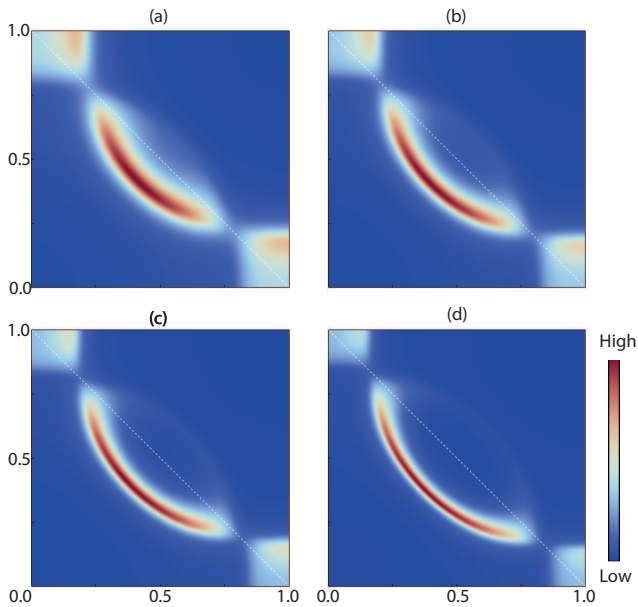}
\caption{(Color online) The spectral function of the low energy theory for each hole doping level, (a) $x=0.05$, (b) $x=0.08$, (c) $x=0.12$, and (d) $x=0.18$. In the present figures, an incoherent background is removed and the intensities of the spectral function are normalized for the first quadrant of the full Brillouin zone. }\label{fig:qp_peaks}
\end{figure}

Consider the third dispersion, Fig.~\ref{fg:disp}(c) in which
$\varphi$ is the mediator of a composite charge $e$ state.  This
corresponds to a self-energy given by 
$\Sigma_-$.  The break in the dispersion just above the chemical
potential is not followed by a re-entrant crossing at a higher
momentum.  Such a re-entrant crossing would give rise to a closed
Fermi surface.  It is the presence of the additional propagating
degree of freedom which thwarts this re-entrance. In addition, there
is no crossing at $(\pi,0)$, but a broad incoherent feature indicative
of the pseudogap.  Since the break-up of the bound state results in a
band crossing near the $(\pi,0)$ region,  the root cause of the
pseudogap is the bound state formed between the bosonic field, $\varphi$ and
a hole as we have advocated previously\cite{charge2e3}. Consequently,
the pseudogap problem is one of confinement. The corresponding Fermi surfaces are shown in
Fig.~\ref{fig:qp_peaks}. The arc-like structure is evident.  The line
of zeros is given by the divergence of the self-energy and hence it is
doublon-holon binding that is responsible for killing the intensity on
the back side of the arc.  The Fermi surfaces evolve smoothly for the doping levels shown from $x=0.05$ to $x=0.18$.
Note also the broad feature at the zone boundary. While it is tempting to interpret the broad peak near the antinodal region as an electron pocket, the lack of coherent excitations makes this view untenable. 

\section{Final Remarks}

The key point this work demonstrates is that two types of charge carriers go into forming Fermi arcs. The projected electrons are present in any low-energy
reduction of the Hubbard model and create the spectral weight on the
high-intensity side of the arc.  The zeros correspond to composite
excitations which are present as a result of dynamical spectral weight
transfer and hence are present only if the UHB is retained or treated appropriately.  Such composite excitations enter the self-energy through
the particle-hole channel, as the relevant scattering process is that
of a fermion from the composite excitation.  Both of these features leading to an effective two-fluid model\cite{charge2e3,charge2e4}
are present within a Wilsonian reduction of the high-energy scale in
the Hubbard model.  The treatment we have derived here should be valid
as long as the UHB provides a relevant perturbation to the physics of
the LHB. Hence, it cannot describe the crossover to the Fermi liquid
regime in which $\varphi$ is unbound. Experimentally, a decoupling of
the UHB from the LHB appears to take place around $x\approx
0.25$\cite{mottcollapse}. Accompanying the collapse is a transition
from a small Fermi surface scaling with $x$ to a
large one with effective area $1-x$.  The precise nature of this
transition will be the subject of a future study.  However, a
prediction of this work is that the in the pseudogap regime, the
volume of the Fermi arc region should be given by $x^\prime$ rather
than $x$.  This follows from the fact that the number of particle-like
excitations that minimally couple to the electromagnetic gauge is less than
$1-x^\prime<1-x$.  Hence, the hole Fermi surface should be given by
$x^\prime$. High precision ARPES measurements can be employed to
verify this result.  

Since our scheme of two type of charge carriers, one giving rise to zeros and the other to poles, seems quite general, it is tempting to rewrite the IR
theory in terms of the composite and projected excitations.  This
would require integrating in an additional field for the composite,
$f_{i\sigma}$ degree of freedom.  The composite fermion is not
canonical, however, and treating it as such would destroy the key feature
leading to a suppression of the spectral on the back-side of the arc. Thus far, we have found no consistent way of doing this.
Hence, an open problem remains precisely how the new composite
excitation should be treated.  But that it is present in any standard
model of Fermi arcs is not in doubt.

\acknowledgements
We thank P.~Johnson for sharing his data with us and extensive
discussions and P.~Ghaemi for his comments on this manuscript and for pointing out Ref.~41.  We also
acknowledge financial support from the NSF-DMR-0940992 and
DMR-1104909 and the Center for Emergent superconductivity, a DOE
Energy Frontier Research Center, Award Number DE-AC0298CH1088.
\bibliographystyle{apsrev4-1}
\bibliography{FA_2eb-1}	

\end{document}